\documentclass[11pt]{article}
\usepackage{epsfig}
\usepackage{amssymb}

\def\bea{\begin{eqnarray}}
\def\eea{\end{eqnarray}}

\def\units{{\ {\rm cm}^{-2}{\rm s}^{-1}}}
\def\srunits{{\ {\rm cm}^{-2}{\rm s}^{-1}{\rm sr}^{-1}}}
\def\svunits{{\ {\rm cm}^{3}{\rm s}^{-1}}}
\def\diffunits{{\ {\rm cm}^{-2}{\rm s}^{-1}{\rm GeV}^{-1}}}
\def\sv{\langle\sigma v\rangle}

\newcommand\prd[3]   
        {{Phys.\ Rev.\ }{\bf D #1} (#2) #3}
\newcommand\prl[3]   
        {{Phys.\ Rev.\ Lett.\ }{\bf #1} (#2) #3}
\newcommand\plb[3]   
        {{Phys.\ Lett.\ }{\bf B #1} (#2) #3}
\newcommand\npb[3]    
        {{Nucl.\ Phys.\ }{\bf B #1} (#2) #3}
\newcommand\app[3]   
        {{Astropart.\ Phys.\ }{\bf #1} (#2) #3}
\newcommand\jhep[3]  
        {{J. High Energy Phys.\ }{\bf #1} (#2) #3}
\newcommand\epjc[3]  
        {{Eur.\ Phys.\ J. }{\bf C #1} (#2) #3}
\newcommand\npps[3]  
        {{Nucl.\ Phys.\ }{\bf #1} {\it(Proc.\ Suppl.)} (#2) #3}
\newcommand\jcap[3]  
        {{JCAP\ }{\bf #1} (#2) #3}

\def\sss{\scriptscriptstyle}

\begin{document}
\begin{titlepage}
\pagestyle{empty}
\baselineskip=21pt
\vspace*{2cm}
\begin{center}
{\LARGE\sf
\mbox{Dark Matter and the CACTUS Gamma-Ray Excess from Draco}}\\[0.4cm]
\end{center}
\begin{center}
\vskip 0.6in
{\Large\sf Stefano~Profumo and Marc Kamionkowski}\\
\vskip 0.2in
{\it {California Institute of Technology, Pasadena, CA 91125, USA}}\\
{E-mail: {\tt profumo@caltech.edu, kamion@tapir.caltech.edu}}\\
\vskip 0.4in
{\bf Abstract}
\end{center}
\baselineskip=18pt \noindent

\noindent The CACTUS atmospheric Cherenkov telescope collaboration
recently reported a gamma-ray excess from
the Draco dwarf spheroidal galaxy. Draco features a very low gas
content and a large mass-to-light ratio, suggesting as a possible
explanation annihilation of weakly interacting massive
particles (WIMPs) in the Draco dark-matter
halo.  We show that with improved angular resolution, future
measurements can determine whether the halo is cored or cuspy,
as well as its scale radius.  We find the relevant
WIMP masses and annihilation cross sections and show that
supersymmetric models can account for the required gamma-ray flux.
The annihilation cross section range is found to be not compatible with a standard thermal relic dark-matter production. We compute for these supersymmetric models the resulting Draco
gamma-ray flux in the GLAST energy range and the
rates for direct neutralino detection and for the flux of
neutrinos from neutralino annihilation in the Sun.
We also discuss the possibility that the bulk of the signal
detected by CACTUS comes from direct WIMP annihilation to two
photons and point out that a decaying-dark-matter scenario for
Draco is not compatible with the gamma-ray flux from the
Galactic center and in the diffuse gamma-ray background.

\vfill
\end{titlepage}
\baselineskip=18pt

\noindent\rule\textwidth{.1pt}
\tableofcontents
\vspace*{0.5cm}
\noindent\rule\textwidth{.1pt}

\vspace*{0.5cm}


\section{Introduction}

Despite compelling indirect astrophysical and cosmological
evidence, the fundamental nature of non-baryonic dark matter
remains elusive (see Refs.~\cite{dmreviews1,dmreviews2} for recent reviews). 
Upgrades of direct detectors looking for scattering of
dark-matter particles from nuclei, future neutrino telescopes, and
space-based antimatter and gamma-ray detectors will dramatically
enhance the possibility to discover dark-matter
particles in the Galactic halo. Given our complete ignorance on
the nature of dark matter, any anomalous experimental
result that might point to dark matter should
be carefully analyzed, and possibly cross-correlated both with
other observational constraints and with existing theoretical
models.

Among indirect-detection techniques that seek products
of dark-matter annihilations, those searching for neutrinos and photons
play a special role. Unlike electrically charged
particles, like antiprotons or positrons, which diffuse in
the Galactic magnetic field, the arrival direction of neutrinos
and photons points to where the dark-matter
annihilation took place. In particular, the forthcoming launch of
the Gamma Ray Large Area Space Telescope (GLAST) \cite{glast} and
the rapidly developing field of ground based Atmospheric Cherenkov
Telescopes (ACTs) \cite{ACTreview} make the search for energetic photons
from dark-matter annihilation a particularly promising and
exciting endeavor.

Since the rate per unit volume for dark-matter annihilation
depends on the square of the dark-matter density, there
may be great advantage to seeking astrophysical locations where
the dark-matter density is believed to be high.  The center of
the Milky Way has been viewed as a promising target as it might
host a dark-matter spike and/or because the Galactic halo could
feature a steep cusp towards its center \cite{pierobuck}. 
On the other
hand, diffuse gamma-ray backgrounds produced by the spallation of cosmic rays
on interstellar gas, and the uncertainties on the
distribution of molecular hydrogen in the galactic ridge, may
blur the dark-matter-induced signal (but for a different
viewpoint see Ref.~\cite{deboer}).  The presence
of other gamma-ray emitters in the central region of our Galaxy,
including those associated with the supermassive central black hole
and the supernova remnant Sgr A$^*$, however, also make it difficult to
distinguish a putative dark-matter signal and its astrophysical
background \cite{galcenter,profumo}.

It has therefore been suggested that one should instead seek
dark-matter-induced gamma rays from other sources that feature
large dark-matter densities but that remain devoid of background
sources.  In particular, several extragalactic sources have been
considered \cite{tyler,baltz,pieri,extragal,evans}.  Among these, dwarf spheroidal
(dSph) galaxies are among the most dark-matter dominated
systems, featuring mass-to-light ratios as high as 
$M/L \sim250\, M_\odot/L_\odot$ \cite{dsphrev}.  Moreover, at
least four dSph galaxies with very large $M/L$ lie within 100
kpc from the Milky Way center.

The dark-matter-induced photon flux depends on the
dark-matter distribution, and particularly on its inner
structure, which is unfortunately poorly known in the case of
local dSphs. Ref.~\cite{baltz} modeled the dark-matter halos of
the four nearby dSph, Carina, Draco, Ursa Minor, and Sextans, with
King profiles. Even with optimistic assumptions on the
background-rejection capabilities of ACTs, they found
that the gamma-ray flux expected from supersymmetric models from
local dSphs is typically at least three orders of magnitude
below the background.
They find that if the dark-matter distribution is clumpy, the signal
can be boosted by at most a factor of 40, which would still be
insufficient. A re-examination of the Draco dSph galaxy in
Ref. \cite{tyler} arrived at somewhat different conclusions. This
work showed that, assuming a steeply cusped
isothermal dark-matter halo, a large portion of the
supersymmetric parameter space could produce a signal visible at both
forthcoming ACTs and at GLAST.  This work also pointed out that,
depending on the magnetic-field strength inside Draco, the
radio-continuum limits for Draco obtained in
Ref.~\cite{dracoradio} 
could also rule out sizeable portions of the same parameter space.
Ref.~\cite{pieri} conducted an extensive
analysis of the 44 nearest
galaxies in the Local Group, pointing out other promising
extragalactic candidates, such as M31. A systematic comparison
between the case of the Galactic center and that of various nearby
dSph galaxies (Sagittarius, Draco, and Canis Major) was carried out
in Ref.~\cite{evans}, where a wide array of halo
profiles was also employed. Depending on the latter,
Ref.~\cite{evans} concludes that if dark matter is
supersymmetric, dSph galaxies may indeed produce a detectable
signal in ACTs and at GLAST.

Recently, a gamma-ray excess from the direction of Draco has
been detected by the Solar 2
Heliostat Array CACTUS, located in Barstow, California \cite{cactus}.
Although the robustness of the signal has still to be tested, the
possibility of ascribing the excess over the off-source
background to dark-matter annihilation in Draco's halo is certainly
intriguing, as Draco is a dark-matter dominated system (see
Sec.~\ref{sec:dracodm}) that is not expected to host any other
significant gamma-ray source \cite{tyler,nogammatyler}. This
possibility was recently envisaged in Ref.~\cite{hooper}. In the
present analysis, we study the impact on the gamma-ray flux of
various, astrophysically motivated, halo models
\cite{mash}. We point out that if the angular resolution of the
experiment can be improved and understood, then measurement of
the angular distribution of the gamma-ray excess 
can discriminate between cuspy (e.g., the Navarro-Frenk-White
profile \cite{nfw}) and cored profiles and determine the halo
scale radius (Sec.~\ref{sec:dracoang}). In
the absence of a full analysis of the raw counts
reported by CACTUS \cite{cactus}, the estimate of the flux of
photons to be attributed to dark-matter annihilation
appears particularly critical. Here, we conservatively estimate
the putative gamma-ray flux detected by CACTUS \cite{cactus}
bracketing it between an upper estimate where all the CACTUS
excess counts over the background are attributed to the
dark-matter signal (following the approach of
Ref.~\cite{hooper}), and a lower estimate where only the counts
in the innermost
angular region are supposed to originate from dark-matter
annihilations. We then scrutinize the dark-matter interpretation
of the
CACTUS excess from Draco on model-independent grounds on the
planes defined by the particle mass and
annihilation cross section
(Sec.~\ref{sec:dracopdm}). We specialize to the particular case of
supersymmetric dark matter in Sec.~\ref{sec:susy}, where we also
make predictions for the expected detection rates for models
consistent with the Draco excess with other dark-matter searches.
We consider in Sec.~\ref{sec:line} the possibility
that the bulk of the excess photons detected by CACTUS come
from the monochromatic $\gamma\gamma$ line.  In
Sec.~\ref{sec:decay}, we show that the signal cannot come from
the decay of a very long-lived dark species. We finally draw our
conclusions in Sec.~\ref{sec:conclusions}.

\section{The dark-matter halo of Draco}\label{sec:dracodm}

Draco was the first dSph galaxy to show evidence of a large
dark-matter content. A total mass much larger than the amount of
visible matter was inferred in Ref.~\cite{aaronson} from the
measurement of four carbon stars in Draco as early as 1983. A few
years later, the assessment of the velocities of 15 more stars
allowed Ref.~\cite{aaronole} to compute a mass-to-light ratio
larger than 50 $M_\odot/L_\odot$, a result subsequently confirmed
by a much larger sample of 91 Draco stars in Ref.~\cite{armand}
and 17 stars in Ref.~\cite{hargre}. More recently, the existence
of an extended dark-matter halo was inferred in
Refs.~\cite{klenya1,klenya2} from a sample of radial velocities
of 159 giant stars out to large projected radii; the
previously assumed hypothesis that mass follows light was shown to
be inconsistent at the 2.5-$\sigma$ confidence level, and tentative
evidence for a nearly isothermal dark-matter distribution was provided.
A new re-evaluation of the Draco dark-matter distribution,
based on new data \cite{wilkinson2004} and on two-component
high-resolution N-body simulations, together with cosmological
predictions for the properties of dark-matter halos, lead
Ref.~\cite{mash} to a few important conclusions, relevant for the
present discussion. First, it was pointed out that both a cored
and a cuspy profile are compatible with the observational
data on Draco and with the results of numerical
simulations.  Second, the possibility that Draco is a tidal dwarf
(i.e., a virtually unbound stellar stream tidally disrupted in the
Milky Way gravitational potential \cite{kroupa}) was ruled out.
This might very well be in contrast with other local dwarf
galaxies \cite{mash,evans}, for which a smaller amount of data is
available.

\begin{figure}[!t]
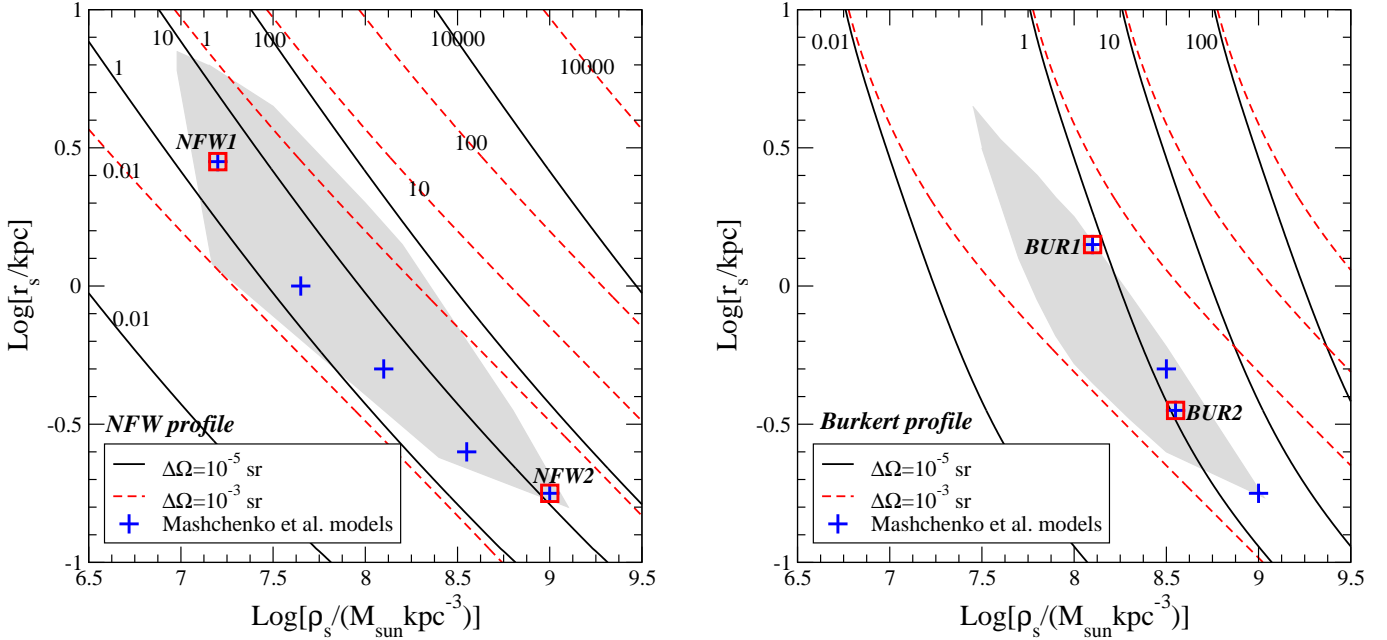

\begin{center}
\hspace*{-0.7cm}\mbox{\epsfig{file=plots/NFW.eps,height=8.5cm}\quad\quad\epsfig{file=plots/BUR.eps,height=8.5cm}}
\end{center}
\caption{\it\small Curves of constant line-of-sight integral of
     the square of the dark-matter density $J_{\Delta\Omega}$ (see
     Eq.~(\protect{\ref{eq:j}}) for the definition of this quantity),
     for $\Delta\Omega=10^{-5}$ (solid black lines) and
     $\Delta\Omega=10^{-3}$ (dashed red lines), on the $(\rho_s,r_s)$
     plane.  For an NFW (left) and for a Burkert (right) dark-matter
     halo, gray regions correspond to cosmologically allowed ranges of
     the two halo parameters, while models indicated with blue crosses
     also give consistent simulated stellar velocity dispersions.
     We highlight with red squares the four benchmark
     halo models of Table~\protect{\ref{tab:hm}}.} \label{fig:hm}
\end{figure}

The analysis of Ref.~\cite{mash} assumes two types of (spherical)
dark-matter halos.  The first one, motivated by $N$-body
simulations, is the well known Navarro-Frenk-White (NFW)
density profile \cite{nfw},
\begin{equation}
     \rho_{\rm\sss NFW}(r)=\frac{\rho_s}{(r/r_s)(1+r/r_s)^2},
\end{equation}
as a function of radius $r$, while the second is the
observationally motivated Burkert
profile \cite{bur},
\begin{equation}
     \rho_{\rm\sss Bur}(r)=\frac{\rho_s}{(1+r/r_s)[1+(r/r_s)^2]}.
\end{equation}
The flux of gamma rays induced by dark-matter annihilation is
clearly very sensitive to the parameters entering the two
profiles; i.e. the scaling density ($\rho_s$) and scale radius ($r_s$).
The $(\rho_s,r_s)$ plane is constrained by observations as well
as cosmological arguments.  We reproduce in Fig.~\ref{fig:hm}
with gray shading the viable 
portions on the $(\rho_s,r_s)$ plane determined in
Ref.~\cite{mash}. The leftmost region is ruled out by the
requirement that the dark-matter halo formed early enough to
allow the subsequent formation of the bulk of its stellar component. The
upper-right limit results from the requirement that the number of
dSph galaxies in the Local group is equal at least to the number
of observed objects. Finally, the viable region is bounded
from below by the condition that Draco's virial radius extends out at
least to the last observed point in its surface-brightness
profile and by consistency with the observed line-of-sight stellar
velocity dispersion. We indicate with blue crosses points
corresponding to models that have $\chi^2<9.5$ between the
modeled and observed line-of-sight velocity-dispersion profiles
according to the analysis of Ref.~\cite{mash}.  
For future convenience, we pick four benchmark halo models,
among those analyzed in Ref.~\cite{mash}, that feature a wide range
of $\rho_s$ and $r_s$; we highlight these models in Fig.~\ref{fig:hm}
with red squares. We collect the scaling densities and radii, as
well as the values of $J_{-3}$ and $J_{-5}$ for the four
benchmark models, in Table~\ref{tab:hm}. 

The observed integral flux of gamma rays with energies
$E>E_\gamma$ from dark-matter annihilation per unit solid angle is
\begin{equation}
     \phi^{E_\gamma}\equiv\frac{N_\gamma\langle\sigma
     v\rangle}{4\pi m_{\rm DM}^2} J_{\Delta\Omega},
\end{equation}
and depends on the particle mass $m_{\rm DM}$, on
its annihilation cross section $\langle\sigma v\rangle$, on the
number $N_\gamma$ of photons  with energies $E>E_\gamma$ produced per
annihilation, and on the integral  $J_{\Delta\Omega}$ along the
line of sight of the square of the dark-matter density, averaged
over the solid angle $\Delta\Omega$ of the detector,
\begin{equation}\label{eq:j}
     J_{\Delta\Omega}\equiv\frac{1}{\Delta\Omega}\int_{\Delta\Omega}\int_{\rm
     l.o.s.}\rho_{\rm\sss DM}^2[r(s)]{\rm
     d}s=\frac{2\pi}{\Delta\Omega}\int_0^{\theta_{\rm max}}{\rm
     d}\theta\sin\theta\int_{s_{\rm min}}^{s_{\rm max}}{\rm d}s\
     \rho_{\rm\sss DM}^2\left(\sqrt{s^2+s_0^2-2ss_0\cos\theta}\right),
\end{equation}
where
\begin{eqnarray}
     &&\theta_{\rm max}\equiv {\rm
     ArcCos}\left(1-\frac{\Delta\Omega}{2\pi}\right),\\ 
     &&s_{\rm min,\ max}\equiv
     s_0\cos\theta\pm\sqrt{r_t^2+s_0^2\sin^2\theta},
\end{eqnarray}
$s_0=75.8\pm0.7\pm5.4$ kpc being Draco's heliocentric distance,
and $r_t$ its tidal radius. The latter was evaluated, following the
approach of Ref.~\cite{evans}, according to the Roche criterion
(the radius such that the average mass in the dSph is equal to the
average interior mass in the Milky Way halo); different tidal
radii correspond to different Milky Way (and, though more weakly,
 to different Draco) dark-matter halos, ranging within less than one order of
magnitude.  However, since $J_{\Delta\Omega}$ depends quite weakly on $r_t\gg
r_s$, we resorted to an isothermal profile for the Milky Way,
which typically gives $r_t\approx7$ kpc.

\begin{table}[!t]
\begin{center}
\begin{tabular}{|c|c|c|c|c|}\hline
{\bf Model}&$\log[\rho_s/(M_\odot\ {\rm kpc}^{-3})]$&$\log [r_s/{\rm kpc}]$&$J_{-3}$&$J_{-5}$\\
\hline
{\bf NFW1}&7.20&0.45&0.12&3.00\\
{\bf NFW2}&9.00&-0.75&0.17&13.0\\
{\bf BUR1}&8.10&0.15&0.26&0.75\\
{\bf BUR2}&8.55&-0.45&0.05&1.13\\
\hline
\end{tabular}
\end{center}
\caption{\it\small Input parameters and line-of-sight integrals
     for four benchmark Draco halo models from
     Ref.~\protect{\cite{mash}}.}\label{tab:hm}
\end{table}

We show in Fig.~\ref{fig:hm} iso-level curves for the quantities
$J_{-3}\equiv J_{\Delta\Omega=10^{-3}\ {\rm sr}}$ and $J_{-5}$,
relevant, respectively, for the full one-degree angular region where
the gamma-ray excess from Draco was observed and for the innermost
0.1-degree angular region, where the largest dark-matter-induced
gamma-ray flux is expected, on the $(\rho_s,r_s)$ plane, for the
NFW profile (left) and for the Burkert profile (right). We
define here and the remainder of the paper the quantities
$J_{\Delta\Omega}$ in units
of $10^{23}\ {\rm GeV}^2\ {\rm cm}^{-5}$. We note that (1) the
observationally and cosmologically consistent range for
$J_{-5}$ spans roughly one order of magnitude in the case of a cored
profile and 2 orders of magnitude for a cuspy profile.  (2)
The ratio $J_{-3}/J_{-5}\approx1$ grows significantly
with $r_s$ for cored profiles, while
$J_{-3}/J_{-5}\approx10\div100$ for cuspy profiles, with smaller
values corresponding to larger scale radii. We find 
as a consistent range for $J_{-5}$ (again in units of $10^{23}\
{\rm GeV}^2\ {\rm cm}^{-5}$),
\begin{eqnarray}
     \nonumber 0.11\lesssim J_{-5} \lesssim 2.96, & & {\rm Burkert\ profile},\\
     \nonumber 0.35\lesssim J_{-5} \lesssim 64.3, & & {\rm NFW\ profile}.
\end{eqnarray}

\section{The CACTUS excess and dark matter}\label{sec:dracosig}

The CACTUS ACT observed a gamma-ray excess over
background from an angular region extending approximately 1 degree
around the direction of the Draco dwarf spheroidal galaxy
\cite{cactus}. The CACTUS energy threshold is around 50 GeV, and
no statistically significant excess for energies greater than 150 GeV was reported.

Given several potential issues concerning the effect of the
integrated starlight from the background and from Draco's stars, the
noise-reduction procedures, and the intrinsic background related
to
misidentified electromagnetic showers induced by hadrons and
electrons, it is extremely difficult to reliably estimate the
{\em signal} gamma-ray flux from Draco.  Some portion of the
excess counts over the background reported by CACTUS will
presumably not be related to the putative dark-matter signal. It
seems reasonable to
assume that the gamma-ray flux is bracketed between two extremes:
as a conservative upper limit, one can consider the overall 
{\em excess} counts detected from the angular region of 1 degree
centered around Draco above the average background measured
outside Draco, which corresponds to approximately 30,000
photons, detected with an effective 
area of the order of $5\times 10^4\ {\rm m^2}$ in 7 hours of
observation \cite{cactus}.  (This procedure was applied in
the recent analysis of Ref.~\cite{hooper}.)  

As an alternative, we can proceed as if the experiment had an
angular resolution of 0.1 degrees.  Currently, the angular
resolution toward Draco is not yet well understood by the CACTUS
collaboration.  A point-source resolution of 0.3 degrees was
obtained toward the Crab nebula, but the resolution toward Draco
may be even poorer.  It is thus premature to attribute the 1
degree spread in the Draco excess to a source of a 1-degree
spatial extent; it may well still be consistent with a point
source.  Still, to illustrate the possibilities with future
measurements with better resolution, we proceed with our
theoretical investigation as if the experiment had such a
resolution.  In this illustrative exercise, one can then
suppose that the signal comes only from the innermost 0.1 degree
angular region around the center of Draco, and that the rest of
the excess is due to spurious effects.  This second procedure is
found to be equivalent, in the
estimate of the signal flux, to requiring that in the innermost
region of Draco, which should contain most of the
dark-matter-induced gamma-ray flux, the ratio of the signal over 
the estimated off-source background for an ACT like CACTUS reproduces the observed relative excess size ($\approx20\%$) \cite{cactus}.

An angular region $\Delta\Omega=10^{-3}$ sr,
corresponding to an angular radius of approximately 1 degree,
gives a signal flux (above the CACTUS energy threshold of 50 GeV)
of $\phi^{50}_{-3}\equiv\phi_\gamma(E_\gamma>50\ {\rm
GeV})\approx2.4\div3.4\times 10^{-9}\units$, where the lower value
corresponds to an effective area $A_{\rm eff}=5\times 10^4\ {\rm
m^2}$, while the
upper value to an average effective area which takes into account
the energy dependence over the interval $50<E_\gamma/{\rm GeV
}<250$ \cite{cactus,hooper}. 

In the second approach, based on the innermost 0.1-degree
angular region to which an ACT may ultimately be sensitive
(corresponding to $\Delta\Omega\approx 10^{-5}$ sr), the main
sources of background correspond to misidentified gamma-like
hadronic showers and cosmic-ray electrons. The diffuse gamma-ray
background should instead not contribute significantly, but we
will take it into account as well. We use the following estimates
for the ACT background \cite{pierobuck}:
\begin{eqnarray}
     \frac{{\rm d}N_{\rm had}}{{\rm d}\Omega}(E>E_0)&=&6.1\times
     10^{-3}\epsilon_{\rm had}\left(\frac{E_0}{1\ {\rm
     GeV}}\right)^{-1.7}\srunits,\label{eq:had}\\ 
     \frac{{\rm d}N_{\rm el}}{{\rm d}\Omega}(E>E_0)&=&3.0\times
     10^{-2}\left(\frac{E_0}{1\ {\rm
     GeV}}\right)^{-2.3}\srunits,\label{eq:el}\\ 
     \frac{{\rm d}N_{\rm diff}}{{\rm
     d}\Omega}(E>E_0)&=&6.7\times 10^{-7}\left(\frac{E_0}{1\
     {\rm GeV}}\right)^{-1.1}\srunits,\label{eq:diff} 
\end{eqnarray}
where $\epsilon_{\rm had}\lesssim 1$ parameterizes the efficiency
of hadronic rejection, and where we took for the diffuse
gamma-ray background the most conservative differential spectral
index (corresponding to the extragalactic gamma-ray background) and normalized the flux to the value of the diffuse emission above
1 GeV from EGRET \cite{egret} in the direction of Draco
($l=86.4^\circ$, $b=34.7^\circ$).  Although the Galactic diffuse
background typically exceeds the extragalactic background at EGRET
energies, the two spectra are such that the extragalactic
background should dominate above 50 GeV.  We therefore derive
an overall background level, using an angular acceptance of
$10^{-5}$sr appropriate for ACTs under our assumptions, of
\begin{eqnarray}
     \phi^{50}_{\rm bckg}\equiv\phi_{\rm bckg}(E>50\ {\rm
     GeV})&\simeq&1.2\times10^{-10}\units,\\ 
     \phi^{150}_{\rm bckg}\equiv\phi_{\rm bckg}(E>150\ {\rm
     GeV})&\simeq&1.5\times10^{-11}\units.
\end{eqnarray}

Comparing the signal outside Draco and towards its center, the
observed excess is around 20\% of the background, which gives a
putative gamma-ray excess
$\phi_{-5}^{50}\approx2.4\times10^{-11}\units$. We let here the
signal vary within one order of magnitude around that central
value, to account for the uncertainties in the background
estimation and in the actual size of the claimed excess. We
therefore consider a signal range, in this approach,
$0.06\lesssim\phi_{-5}^{50}/\phi^{50}_{\rm bckg}\lesssim0.6$.
As mentioned above, this second conservative estimate of the
signal is consistent with the actual number of counts reported
by CACTUS within an angular radius of around 0.1 degrees.

No excess flux has been observed from Draco above 150 GeV. This
leads to a further constraint on the dark-matter interpretation.
We consider two putative upper limits: the strongest comes from the
requirement that in the central bins the signal is less than 5\%
of the ACT backgrounds, and reads
$\phi_{-5}^{150}\lesssim 7.5\times10^{-13}\units$; the most
conservative requirement is instead that the signal flux does
not exceed the Poisson fluctuation of the actual number of
counts above the energy threshold of 150 GeV; this gives a limit
on the signal flux of $\phi_{-5}^{150}\lesssim
12.6\times10^{-12}\units$.

In the following, we will refer to the CACTUS signal as a photon
flux in the range
\begin{equation}\label{eq:dracosig}
     \phi_{-3}^{50}\lesssim 3.4\times10^{-9}\units\quad
     \phi_{-5}^{50}\gtrsim 7.2\times10^{-12}\units \quad
     \phi_{-5}^{150}\lesssim 0.75\div12.6\times10^{-12}\units.
\end{equation}
Again, the $\Delta\Omega=10^{-5}$ numbers do not describe the
current Draco data; rather, they describe results of a
hypothetical experiment with 0.1 degree resolution that look
like the current CACTUS results.

\subsection{The gamma-ray angular distribution}\label{sec:dracoang}

The gamma-ray excess is spread over 1 degree.  However, given
uncertainties in the current CACTUS angular resolution around
Draco, we cannot currently use the observed spread to
discriminate between a source with a 1 degree spatial extent and
a point source.  Still, it is conceivable that forthcoming ACT
measurements may achieve a resolution as good as 0.1 degrees.  In
this Section, we illustrate how future such measurements may
shed light on the halo profile for Draco.  To do so, we proceed
as if the resolution of CACTUS were in fact already 0.1 degrees.

Our second estimate of the flux, depending on the
assumed dark-matter distribution, and particularly on the scale
radius of the dark-matter halo, typically gives a total flux from
the central 1 degree angular region which can be significantly lower
than the total counts reported by CACTUS. Within this approach,
the counts originating from regions outside the center of Draco
would correspond to photons produced by dark-matter annihilation
only if a very large scale radius (of the order of 1 kpc) were
assumed.

\begin{figure}[!t]
\begin{center}
\epsfig{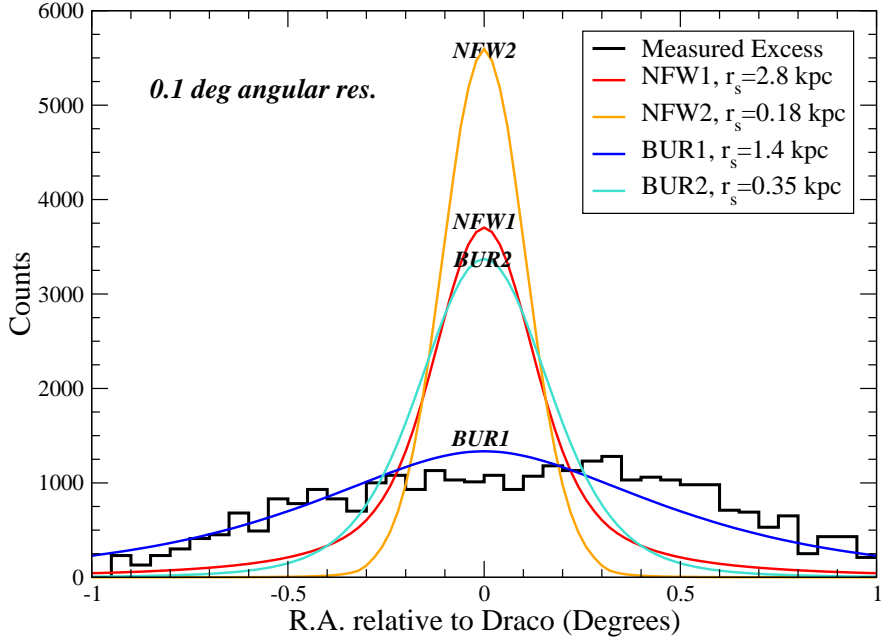}
\end{center}
\caption{\it\small A comparison among theoretical predictions for
the angular distribution of the gamma-ray flux from dark-matter
annihilation and the excess reported by
CACTUS for four different Draco dark-matter halos \cite{mash},
featuring a wide range of scale radii, and giving consistent
$N$-body-simulated stellar-velocity dispersions.  Keep in mind
that the current CACTUS data may have a resolution of roughly
one degree, and so conclusions about the Draco halo cannot yet
be drawn by comparing the current data with the theoretical curves.}
\label{fig:angular}
\end{figure}

What we will now show is that if CACTUS excess is real and if the
Draco halo is cuspy, then an ACT measurement with an 0.1 degree
resolution should see a very peaked angular distribution.  
To do so, we reproduce in
Fig.~\ref{fig:angular} the measured photon 
excess in the 1-degree angular region centered on the
location of Draco.  We compare the
putative excess with the predictions, for the photon-flux angular
distribution, stemming from four different dark-matter halos. We
normalize the photon fluxes for the various halos to the total
number of excess gamma rays reported by the CACTUS preliminary
results. The four profiles were chosen among those quoted in
Table~1 of Ref.~\cite{mash}; two of them are Burkert profiles with
scale radii of 1.4 kpc (BUR1, profile B1 in Ref.~\cite{mash}) and
0.35 kpc (BUR2, profile B2 in Ref.~\cite{mash}), and two of them
are NFW profiles with scale radii of 2.8 kpc (NFW1, profile N1 in
Ref.~\cite{mash}) and 0.18 kpc (NFW2, profile N5 in
Ref.~\cite{mash}). We deduce from Fig.~\ref{fig:angular}
that if the current resolution of CACTUS were as good as 0.1
degrees, then the NFW profiles would produce an excess of gamma rays in the
central bins over what is now seen.  If, however, the current
angular distribution is still observed even with an 0.1 degree
resolution, then cored profiles, with scale radii of order 1
kpc, would be indicated by the data.  We conclude therefore that
improved angular resolution is warranted to understand better
the halo structure.

\subsection{Implications for particle dark matter}\label{sec:dracopdm}

No gamma-ray source in the direction of Draco was identified by
EGRET in its all-sky survey. An analysis was carried
out by the EGRET collaboration in Ref.~\cite{wai}. The absence of a
point-like source from the direction of Draco can be
independently used to draw an upper limit
on the integral flux of photons above a threshold of 1
GeV, following the analysis of Ref.~\cite{lamb}. As in
Ref.~\cite{tyler}, one can take, as the EGRET upper limit on
point-like sources gamma-ray fluxes the flux of
the least significant point-source detection in the EGRET catalog,
corresponding to the Large Magellanic Cloud, which translates into
the requirement
\begin{equation}\label{eq:egret}
     \phi_{-3}^1\equiv\phi(E>1\ {\rm GeV})\lesssim\phi^1_{\rm
     EGRET}\simeq10^{-8}\units.
\end{equation}
This flux agrees with the theoretical estimate for the flux sensitivity of
EGRET to point sources determined in Ref.~\cite{Morselli:2002nw}. 

The actual EGRET data from the direction of Draco were
collected, in Ref.~\cite{wai}, in 7 energy bins, featuring
different angular cuts and different exposures. Every energy bin
is also accompanied by a background estimate. In order to
compare the aforementioned point-source sensitivity of
EGRET with the actual photon count, one needs to pick a
halo profile for Draco and a signal photon
spectrum. We also clustered the four lowest-energy bins ($0.1<E_\gamma/{\rm
GeV}<1$) and the three highest-energy bins ($1<E_\gamma/{\rm
GeV}<10$), since in some bins the background estimate exceeds the
photon count, and in the highest-energy bins the statistics are
very poor. We require that in the two energy intervals the
signal does not exceed the difference between the measured
photon counts and the estimated background.

We find that in general the point-source sensitivity agrees,
within a factor at most 2, with the bound from the {\em
highest}-energy bins, while that from the {\em smallest}-energy
bins is approximately one order of magnitude weaker (and, in
particular, the harder the signal photon spectrum, the weaker
the latter bound). The bound from the highest-energy bins tends to be
stronger than the point-source-sensitivity constraint for (1)
smaller particle masses, (2) for cored rather than
cuspy dark-matter halos, and (3) for softer photon
spectra. Since, however, the point-source-sensitivity
criterion appears to be less dependent on the poor statistics of
the highest-energy bins (consisting of only a few photon
counts), and since the two criteria essentially agree, we
hereafter indicate as the EGRET bound the limit in
Eq.~(\ref{eq:egret}).

The VERITAS collaboration also observed Draco with the Whipple-10m ACT,
reporting a null search for high-energy gamma rays above a
threshold $E_\gamma\simeq400$ GeV, which translates into the
bound \cite{jeter},
\begin{equation}
     \phi_{-5}^{400}\equiv\phi(E>400\ {\rm GeV})\lesssim\phi^{400}_{\rm
     Whipple-10m}\simeq1\times10^{-12}\units.
\end{equation}

The number of gamma rays from the annihilation of a WIMP integrated over an energy $E_{\gamma}$ can be written as
\begin{equation}\label{eq:phigamma}
     N_\gamma(E_\gamma)=\int_{E_{\gamma}}^\infty\ \left(\sum_f\ {\rm
     BR}(\chi\chi\rightarrow f)\frac{{\rm d}N^f_\gamma}{{\rm
     d}E}\right){\rm d}E
\end{equation}
where the symbol $f$ refers to any WIMP-annihilation final
state, yielding a gamma-ray spectral function 
(differential number of photons per WIMP annihilation) ${\rm
d}N^f_\gamma/{\rm d}E$. Different final states give different
spectral functions. We take here as benchmark cases those giving
the hardest and the softest spectra among the final states
that are relevant in the case of supersymmetric dark matter;
i.e., $\tau^+\tau^-$ and $b\bar b$, respectively.

\begin{figure}[!t]
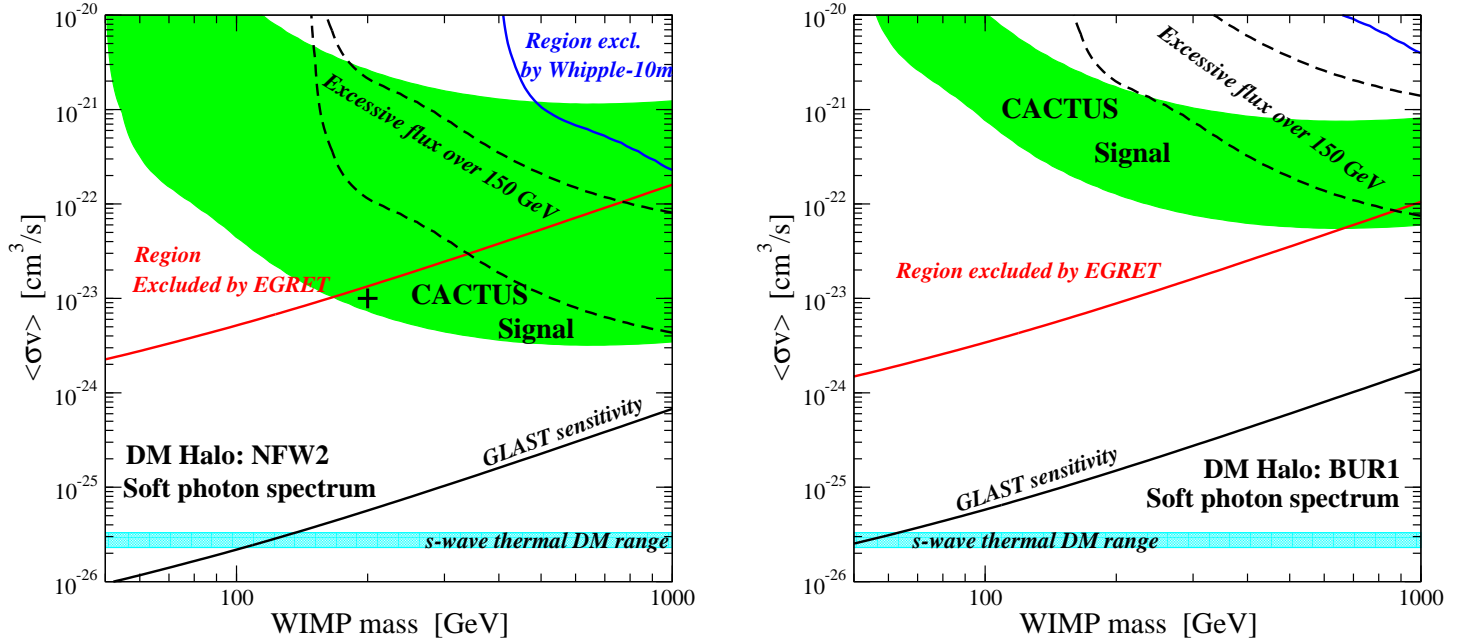

\begin{center}
\hspace*{-0.7cm}\mbox{\epsfig{file=plots/bb_nfw2.eps,height=8.5cm}\quad\quad\epsfig{file=plots/bb_bur1.eps,height=8.5cm}}

\end{center}
\caption{\it\small Regions on the $(m_{\rm\sss WIMP},\sv)$ plane
compatible with the CACTUS excess and with other observations for
 a soft photon spectrum (of the same type as that shown with red
lines in Fig.~\protect{\ref{fig:dnde}}, $b\bar b$ final state).
The two left panels assume the cuspy NFW2 dark-matter halo for
Draco, while those on the right
the cored BUR1 halo (see Table~\ref{tab:hm}).} \label{fig:bb}
\end{figure}

\begin{figure}[!t]
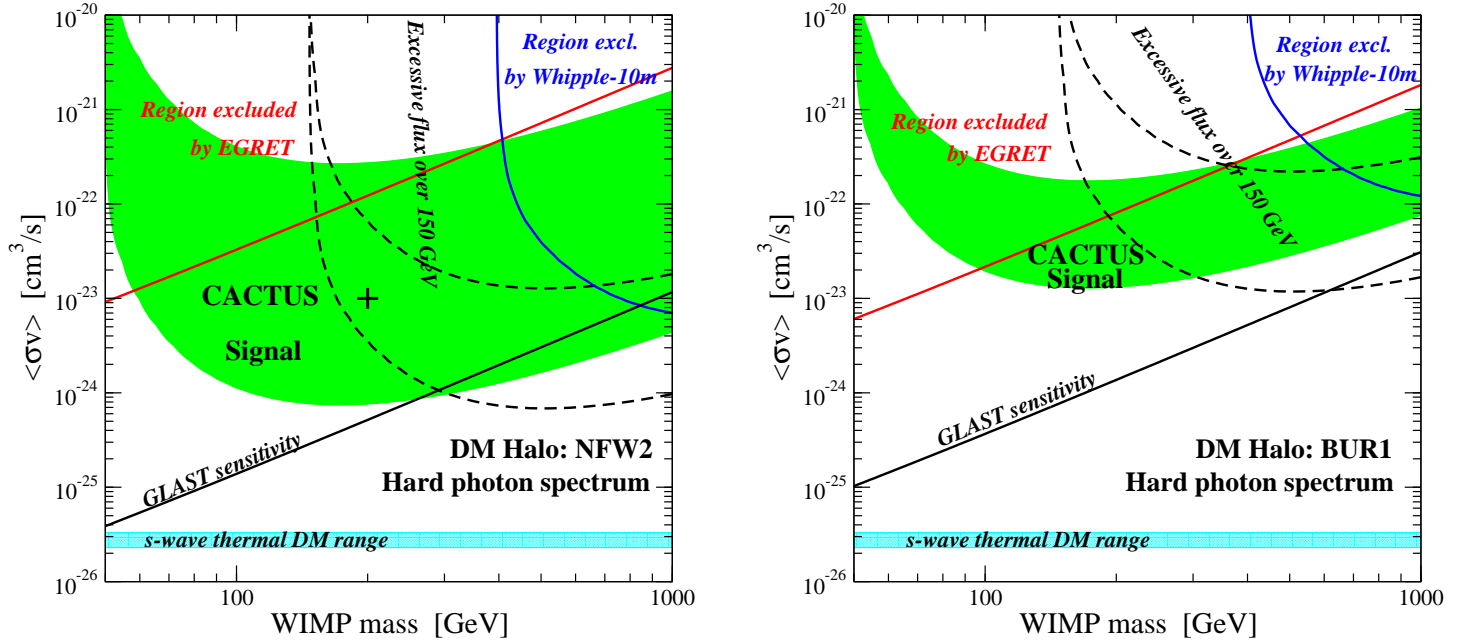

\begin{center}
\hspace*{-0.7cm}\mbox{\epsfig{file=plots/tt_nfw2.eps,height=8.5cm}\quad\quad\epsfig{file=plots/tt_bur1.eps,height=8.5cm}}\\

\end{center}
\caption{\it\small As in Fig.~\ref{fig:bb}, but for a hard photon
spectrum (blue dashed lines in Fig.~\protect{\ref{fig:dnde}},
$\tau^+\tau^-$ final state).} \label{fig:tt}
\end{figure}

We show in Figs.~\ref{fig:bb} and \ref{fig:tt} the region, on the
$(m_{\rm\sss WIMP},\sv)$ plane, compatible with the putative
CACTUS signal quoted in Eq.~(\ref{eq:dracosig}), and the various constraints discussed above, including the two estimates of the limit from the non-observation of a significant excess of gamma rays with energies above 150 GeV.
Fig.~\ref{fig:bb} refers to a soft photon spectrum, corresponding
to the final state $f=b\bar b$, while Fig.~\ref{fig:tt} to a hard
photon spectrum, $f=\tau^+\tau^-$. The two panels to the left to a cuspy
profile (namely, the NFW2 profile, i.e. profile N5 in
Ref.~\cite{mash}), while the two panels to the right employ a cored
dark-matter halo (namely, the BUR1 profile; i.e., profile B3 in
Ref.~\cite{mash}). 

We emphasise how the range of dark-matter annihilation cross sections needed to explain the CACTUS signal is largely above what a naive estimate from the thermal production of relic WIMPs would suggest. Assuming that the pair annihilation rate $\sv$ is energy independent, and therefore that annihilations proceed through $s$-wave processes, the WIMP $\chi$ relic abundance scales as $\Omega_\chi h^2\simeq3\times 10^{-27}\ {\rm cm^3s^{-1}}/\sv$ \cite{dmreviews1}. Requiring that all the CACTUS excess originates from dark-matter annihilations would therefore naively entail thermal relic abundances as low as $\Omega_\chi h^2\sim10^{-5}\div10^{-6}$. For reference, we indicate the $\sv$ range deduced from the previous relation and from the 2-$\sigma$ WMAP estimate of the CDM abundance \cite{Spergel:2003cb} with light blue bands in fig.~\ref{fig:bb} and \ref{fig:tt}. Resonant annihilations, coannihilation processes or thresholds can significantly distort, though, the above outlined estimate: we postpone to sec.~\ref{sec:susy} (see in particular fig.~\ref{fig:jpsi}) a discussion of the compatibility of the CACTUS signal with the thermal production of dark matter for the particular case of neutralinos.

We also include the sensitivity of GLAST, computed assuming a
total exposure of \mbox{$3.2\times 10^{11}\ {\rm cm}^2\ {\rm s}$} and the
diffuse gamma-ray background measured by EGRET,
Eq.~(\ref{eq:diff}), and requiring the strongest among the
following conditions: (1) $N_S/\sqrt{N_B}>5$, $N_S$ and $N_B$ being the
total number of signal and background events, or (2)
$N_S>10$. The angular acceptance was set to maximize
the signal-to-noise ratio (namely, the quantity $J_{\Delta\Omega}\sqrt{\Delta\Omega}$ for a given halo profile) between the minimal angular
resolution and the maximal field of view.

In the case of a soft photon spectrum (Fig.~\ref{fig:bb}),
the EGRET bound rules out large parts of the CACTUS signal,
regardless of the assumed halo profile. With a soft photon spectrum, all models consistent with
the lower estimate of the CACTUS signal will be largely within
the GLAST sensitivity.

With a hard signal photon spectrum (Fig.~\ref{fig:tt}), the bound
from the Whipple-10m observation of Draco is significantly more
effective than with a soft spectrum; again, almost all the
region compatible with the CACTUS signal will be within
GLAST reach. We also remark that with the BUR1 profile
(Fig.~\ref{fig:tt}, right), and with a hard photon spectrum, we
find models that (1) give a total photon flux equal to the
CACTUS counts; (2) give an angular distribution of the photon
counts that is very close to that reported by CACTUS (see
Fig.~\ref{fig:angular}; keep in mind that the theoretical curves reported in Fig.\ref{fig:angular} refer to a much better angular resolution than that of CACTUS); and (3) are consistent with the
conservative criterion on the gamma-ray flux above 150 GeV.

\begin{figure}[!t]
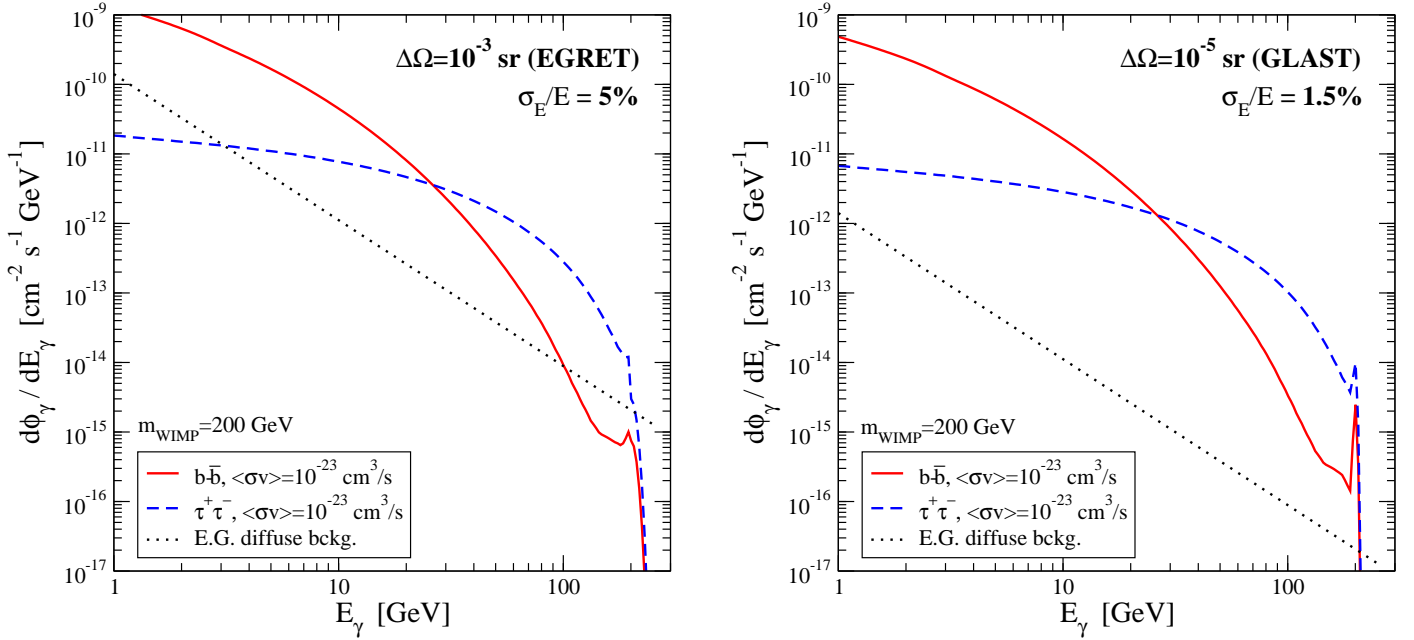

\begin{center}
\hspace*{-0.7cm}\mbox{\epsfig{file=plots/dnde_egret.eps,height=8.5cm}\quad\quad\epsfig{file=plots/dnde_glast.eps,height=8.5cm}}
\end{center}
\caption{\it\small Sample soft (solid red lines, $b\bar b$ final
state) and hard (dashed blue lines, $\tau^+\tau^-$ final state)
photon spectra giving a photon flux compatible with the excess
observed from the direction of Draco by the CACTUS ACT. The WIMP
mass is set to 200 GeV, and the annihilation cross section
times velocity to $\langle\sigma v\rangle=10^{-23}\svunits$. The
dotted lines correspond to an
estimate of the diffuse gamma-ray background, and
the left panel refers to an angular resolution
$\Delta\Omega=10^{-3}$ sr, while the right
panel to an angular resolution $\Delta\Omega=10^{-5}$ sr, with the assumed NFW2 dark-matter halo.} \label{fig:dnde}
\end{figure}

The two black crosses indicate two benchmark points, whose
differential photon spectrum is reproduced in
Fig.~\ref{fig:dnde} for illustrative purposes, together with the expected
diffuse gamma-ray background, for a space-based gamma-ray
telescope with angular resolution $\Delta\Omega=10^{-3}$ sr
(EGRET-like) and $\Delta\Omega=10^{-5}$ sr (GLAST-like). We also
show the monochromatic gamma-ray line, assuming $\langle \sigma
v\rangle_{\gamma\gamma}/\langle \sigma v\rangle_{\rm tot}=3\times
10^{-4}$ (close to the maximal value for supersymmetric dark
matter; see Sec.~\ref{sec:line}),
and different energy resolutions, appropriate for the two
space-based detectors.

\subsection{Supersymmetric dark matter}\label{sec:susy}

Our analysis was based, up to this point, on a model-independent
approach as far as the annihilation cross section and the
final-state branching ratio for dark-matter annihilation are
concerned. In order to predict the rates for
dark-matter detection in other search arenas, such as direct
detectors and ${\rm km}^3$ neutrino telescopes, one needs
to consider specific particle models. To this extent,
we now specialize to supersymmetric dark
matter and consider the minimal $CP$-conserving supersymmetric
extension of the standard model (MSSM), and perform a random
scan over its parameter space. For all models, we impose
constraints from direct supersymmetric-particles searches at
accelerators, rare processes with a sizeable
potential supersymmetric contribution, the lower bound on the mass
of the lightest $CP$-even Higgs boson, and precision electroweak tests. We also
require the lightest supersymmetric particle (LSP) to be the
lightest neutralino. We do not, however, require that the thermal
relic abundance $\Omega_\chi$ of the LSP falls within the CDM
abundance determined within the $\Lambda$CDM paradigm.  We
assume that non-thermal production of neutralinos in the early
Universe \cite{nonth}, or cosmological enhancements of the relic neutralino
density \cite{enh}, brought $\Omega_\chi\simeq\Omega_{\rm CDM}$. We detail in
Table~\ref{tab:scan} the MSSM scan procedure.

\begin{table}[!t]
\begin{center}
\begin{tabular}{|c|c|c|c|c|c|c|c|}\hline
$\mu$ & $m_1$ & $m_2$ & $m_3$ & $m_A$ & $m_{\widetilde S}$ & $A_{\widetilde S_3}$ & $\tan\beta$\\
\hline
$30\div1200$ &  $2\div1200$ &  $50\div1200$ &  $m_{\rm\sss LSP}\div20000$ &  $100\div10m_{\rm\sss LSP}$ &  $(1\div10)m_{\rm\sss LSP}$ &  $(-3\div3)m_{\widetilde S}$ &  $1\div60$\\
\hline
\end{tabular}
\end{center}
\caption{\it\small Ranges of the MSSM parameters used to generate
the models shown in Figs.~\protect{\ref{fig:scan}} and
\protect{\ref{fig:jpsi}}. All masses are in GeV, and
$m_{\rm\sss LSP}\equiv{\rm min}(\mu,m_1,m_2)$. The quantity
$m_{\widetilde S}$
indicates the following scalar masses (which were independently
sampled): $m_{\widetilde Q_{1,3}}$, $m_{\widetilde u_{1,3}}$,
$m_{\widetilde d_{1,3}}$, $m_{\widetilde L_{1,2,3}}$,
$m_{\widetilde e_{1,2,3}}$. To avoid FCNC constraints, we assumed
the squark soft supersymmetry breaking terms of the first two
generations to be equal. $A_{\widetilde S_3}$ stands for the third
generation sfermion trilinear terms: those of the first two
generations were taken to vanish.}\label{tab:scan}
\end{table}
\begin{figure}[!t]
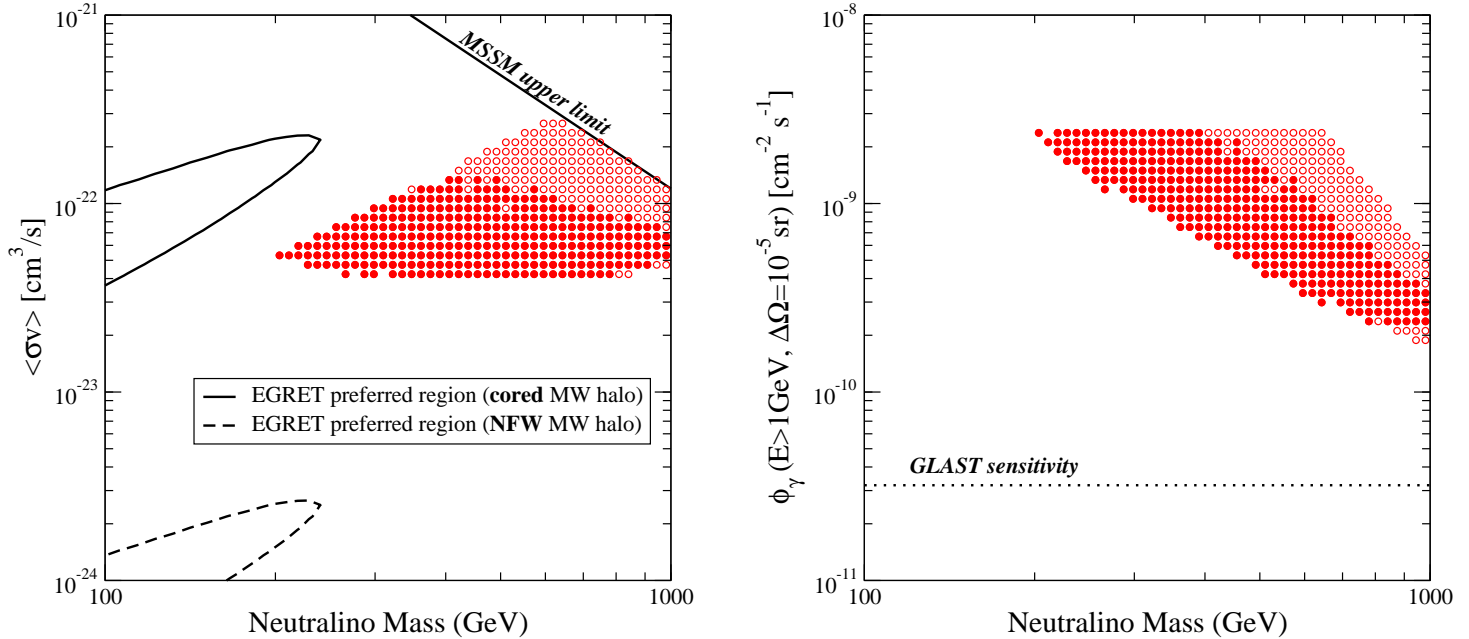

\begin{center}
\hspace*{-0.7cm}\mbox{\epsfig{file=plots/mx_sv.eps,height=8.5cm}\quad\quad\epsfig{file=plots/mx_glast.eps,height=8.5cm}}\\[2cm]

\end{center}
\caption{\it\small We show the results of a scan over the general
MSSM (see Table~\protect{\ref{tab:scan}} for details) where we
looked for supersymmetric models providing a flux of gamma rays
from Draco compatible with the CACTUS excess. The left panel shows the $(m_{\rm
WIMP},\sv)$
plane, while the right panel shows
the expected integral flux of photons at GLAST, with the
corresponding estimated point-source sensitivity in the direction of Draco.}
\label{fig:scan}
\end{figure}

\begin{figure}[!t]
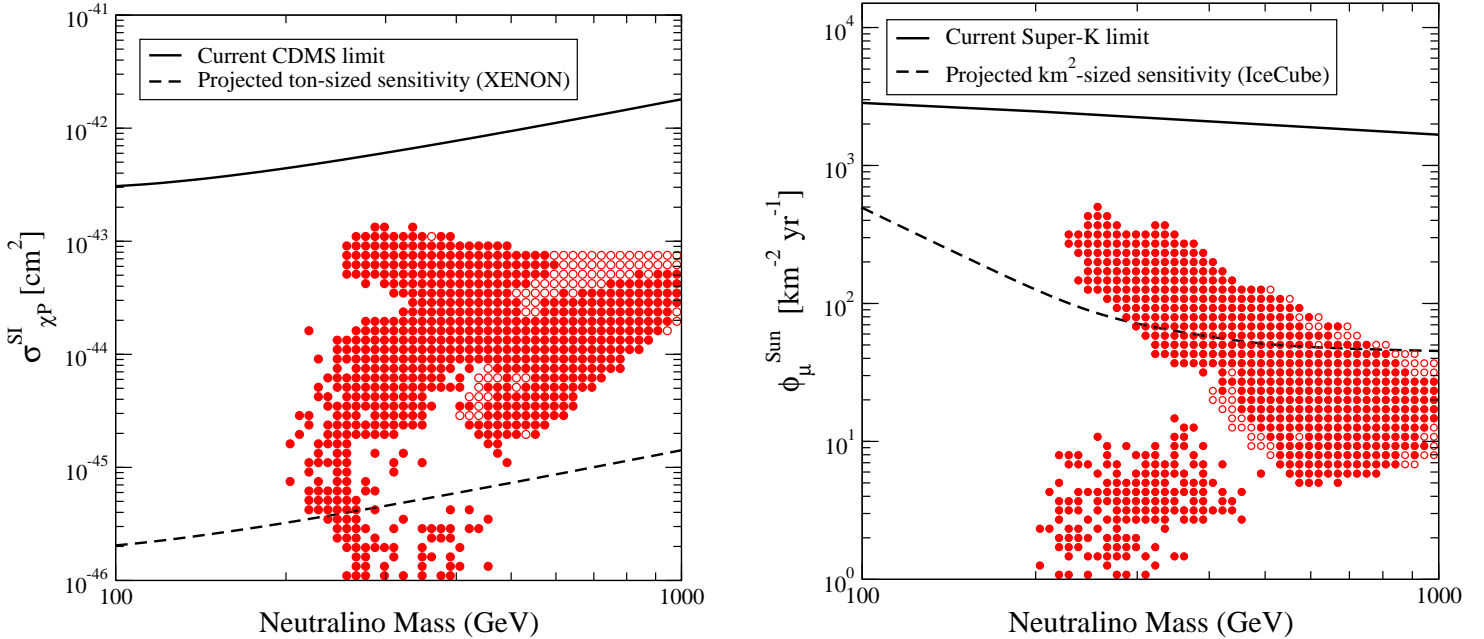

\begin{center}
\hspace*{-0.7cm}\mbox{\epsfig{file=plots/mx_dir.eps,height=8.5cm}\quad\quad\epsfig{file=plots/mx_neusun.eps,height=8.5cm}}\\

\end{center}
\caption{\it\small The neutralino-proton scalar-interaction cross
section and the flux of muons from the Sun induced by neutralino
annihilations, with the corresponding present and future
experimental sensitivities, for the same models as in
Fig.~\protect{\ref{fig:scan}}.} \label{fig:scan2}
\end{figure}

In the upper left panel of Fig.~\ref{fig:scan}, we indicate on the
$(m_\chi,\sv)$ plane supersymmetric models that give a gamma-ray
flux from Draco in the range of Eq.~(\ref{eq:dracosig}).
Filled points are also consistent with the (strongest) requirement of a
sufficiently low gamma-ray flux above 150 GeV. All models are
consistent, instead, with the conservative constraint, from the
Poisson variance of the measured counts, on the
flux above 150 GeV. We employ here the BUR2 profile of
Table~\ref{tab:hm}.
We also indicate the regions that
significantly improve the fit to the gamma-ray flux from the
direction of the Galactic center as measured by EGRET
\cite{egretgc}, for a cored and an NFW Milky Way dark-matter halo.
Intriguingly enough, for a Milky Way halo that is slightly
cuspier than a Burkert (cored) profile (solid black line in Fig.~\ref{fig:scan}, right), the region favored by the EGRET
data can be consistent with the Draco signal, provided the neutralino mass is
not much heavier than 250 GeV. We also indicate the upper limit
on the annihilation cross section in the general MSSM
derived in Ref.~\cite{profumo}, and note that most models fall
very close to the maximal cross sections in the context of
supersymmetric dark matter. Such large cross sections could be
in conflict with the production of other secondary
annihilation products, such as antiprotons, positrons, or
antideuterons \cite{Profumo:2004ty,Baer:2005tw}. However, the
large uncertainties in the modeling of diffusion
processes and nuclear reactions, together with
those connected to the Milky Way dark-matter halo, can
leave the freedom to circumvent those constraints
\cite{Baer:2005tw}. 

We find that over the whole scanned supersymmetric parameter
space, the total flux over the whole 1 degree angular region
features, with the BUR2 profile we employ here, an upper limit
$\phi^{50}_{-3}\lesssim2\times10^{-10}$, with a maximum at a
neutralino mass of 600 GeV. This upper bound indicates that no
supersymmetric models can give, in the present halo setup, 100\%
of the excess counts over the off-source background detected by CACTUS. The
latter could be achieved with a cuspier NFW profile, the scaling
among the fluxes simply being $\phi^{\rm NFW}/\phi^{\rm
BUR}=J^{\rm NFW}/J^{\rm BUR}$ (see
Fig.~\ref{fig:hm}). 

We then carry out an analysis of the prospects for the detection
of the CACTUS-compatible supersymmetric models with other
search avenues. In particular, in the right panel of
Fig.~\ref{fig:scan}, we show the total photon flux expected at
GLAST, integrated over an energy threshold of 1 GeV, and with
$\Delta\Omega=10^{-5}$ sr (which maximizes the signal to noise
with the profile we use here). All CACTUS-compatible
supersymmetric models will be, under the present assumptions,
unambiguously within the sensitivity of GLAST. The upper bound
on the integrated flux of photons stems from the EGRET bound,
Eq.~(\ref{eq:egret}), since $\phi^{\rm GLAST}=10^{-2}\phi^{\rm
EGRET}(J_{-5}/J_{-3})\approx2.3\times10^{-9}\units$.

Fig.~\ref{fig:scan2} collects, again for the CACTUS-compatible
supersymmetric models discussed above, the results for the rates
at future direct-detection experiments (left) and at IceCube (right),
where muons induced by neutrinos from dark-matter annihilation in the Sun
will be sought.
Although not guaranteed, the detection prospects for
supersymmetric models featuring the correct mass and
annihilation cross sections certainly look promising.

\begin{figure}[!t]
\begin{center}
\mbox{\epsfig{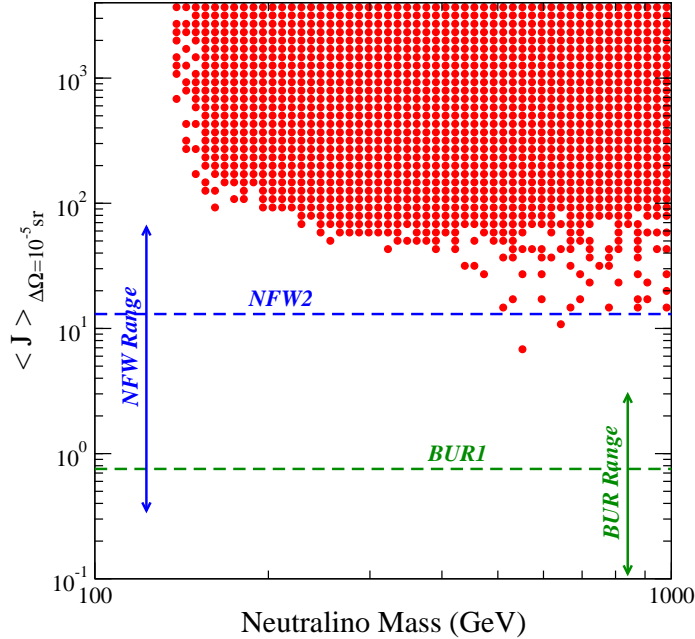}}
\end{center}
\caption{\it\small Scan of the general MSSM (see
Table~\protect{\ref{tab:scan}}) showing the putative values of
$J_{-5}$ needed to explain the CACTUS excess for
supersymmetric models featuring a thermal relic abundance
$\Omega_\chi h^2\simeq\Omega_{\rm CDM}h^2$. The horizontal
lines indicate the values of $J_{-5}$ for a sample of
Draco halo profiles \protect{\cite{evans}}, including the NFW2
and BUR1 profiles of Figs.~\ref{fig:bb} and \ref{fig:tt}. }
\label{fig:jpsi}
\end{figure}

If we require the thermal neutralino relic abundance to agree
with the CDM abundance, then strong constraints are placed to
the range of annihilation cross sections.  One can then ask,
for neutralino models with the correct thermal abundance, 
what values of $J_{-5}$ are required to account for the CACTUS excess?
We answer this question in
Fig.~\ref{fig:jpsi}, where we indicate, for MSSM models
obtained in the same scan outlined in Table~\ref{tab:scan} that
fulfill the condition $\Omega_\chi\simeq\Omega_{\rm CDM}$,
the values of $J_{-5}$ such that
$\phi_{-5}^{50}=2.4\times10^{-11}\units$. We show with vertical lines the values of $J_{-5}$ for the profiles employed in Figs.~\ref{fig:bb} and
\ref{fig:tt}. The vertical arrows indicate the overall ranges for
$J_{-5}$ obtained in the analysis of Sec.~\ref{sec:dracodm}.
Evidently, even the minimal gamma-ray flux compatible
with the CACTUS excess requires an extremely cuspy halo for Draco.
As we pointed out before, such cuspy profiles should be readily identifiable once a better angular resolution is achieved through the analysis of the angular distribution of the photon excess counts. 

\subsection{Monochromatic gamma rays}\label{sec:line}
\begin{figure}[!t]
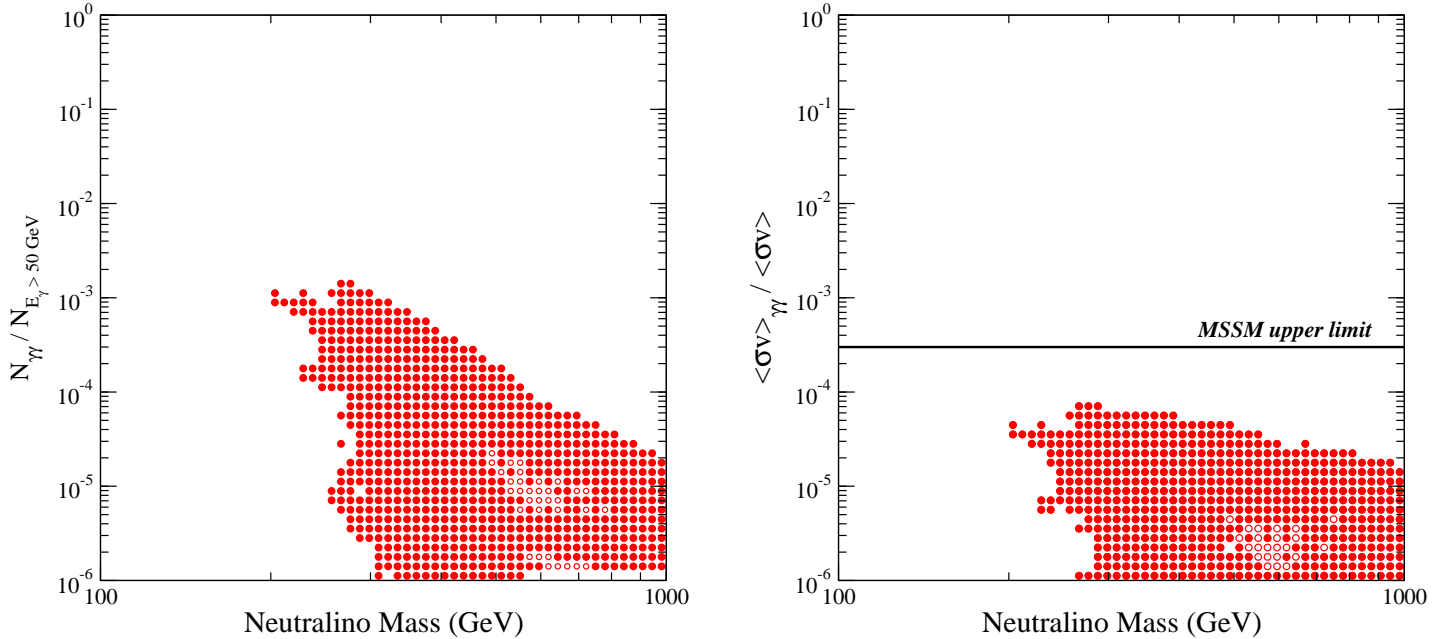

\begin{center}
\hspace*{-0.7cm}\mbox{\epsfig{file=plots/gr_line.eps,height=8.5cm}\quad\quad\epsfig{file=plots/sv_line.eps,height=8.5cm}}
\end{center}
\caption{\it\small Left: The ratio between the number of photons
from direct pair annihilation of neutralinos into two photons and
from the gamma-ray continuum, integrated above 50 GeV, for the
CACTUS-compatible supersymmetric models of
Fig.~\ref{fig:scan}. Right: the ratio of the annihilation
cross section of neutralinos into two photons over the total
neutralino annihilation cross section, for the same ensemble
of models.} \label{fig:linesusy}
\end{figure}

Given the preliminary spectral structure of the CACTUS Draco
gamma-ray excess---i.e. a large
amount of counts in a narrow energy range---it is certainly
worthwhile to investigate whether the 
excess may be due to direct annihilation of dark-matter particles to
photons.  Unfortunately, the poor energy resolution of the
solar-array ACT is not suitable for a prompt discrimination of
this possibility, which, if detected, would constitute a smoking
gun for dark-matter annihilation. A dominantly monochromatic
signal would moreover circumvent the friction with the EGRET
null result.

As a first step, we again consider here supersymmetric dark
matter, and determine whether a scenario
where the bulk of the dark-matter-induced gamma-ray flux
originates from direct neutralino annihilation into two photons is
at all viable. To this extent, we show in
the left panel of Fig.~\ref{fig:linesusy} the relative fraction of
monochromatic photons versus the integrated number of continuum
photons with energies larger than 50 GeV, for supersymmetric
models giving a CACTUS-compatible gamma-ray flux, as determined in
the previous Section. In the right panel, we show the relative
branching fraction for neutralino annihilation into photons.
We conclude that within supersymmetric models, the bulk of the
photon flux always stems from the continuum component, the
monochromatic part contributing less than $\approx0.2\%$ of the photon
counts for CACTUS-compatible supersymmetric models.  As a
byproduct, we derived from the full scan an upper bound on the
branching ratio of neutralino
annihilation into two photons in the general MSSM which reads
\begin{equation}
     r\equiv\langle\sigma v\rangle_{\gamma\gamma}/\langle\sigma
     v\rangle_{\rm tot}\lesssim3\times10^{-4}.
\end{equation}

On more general grounds, outside the supersymmetric
paradigm, one can still hypothesize that the bulk of the CACTUS
signal comes from the monochromatic line, and constrain, through
the EGRET bound, the continuum contribution, and hence the
quantity $r$. We point out that the $Z\gamma$ or the $Zh$ lines would
constitute less favorable scenarios here, since (1) the
monochromatic photon flux would be smaller by a factor 0.5, and
(2) the continuum photons from the $Z$ and $h$ decay would contribute to
the continuum photon yield, strengthening the EGRET bound.

\begin{figure}[!t]
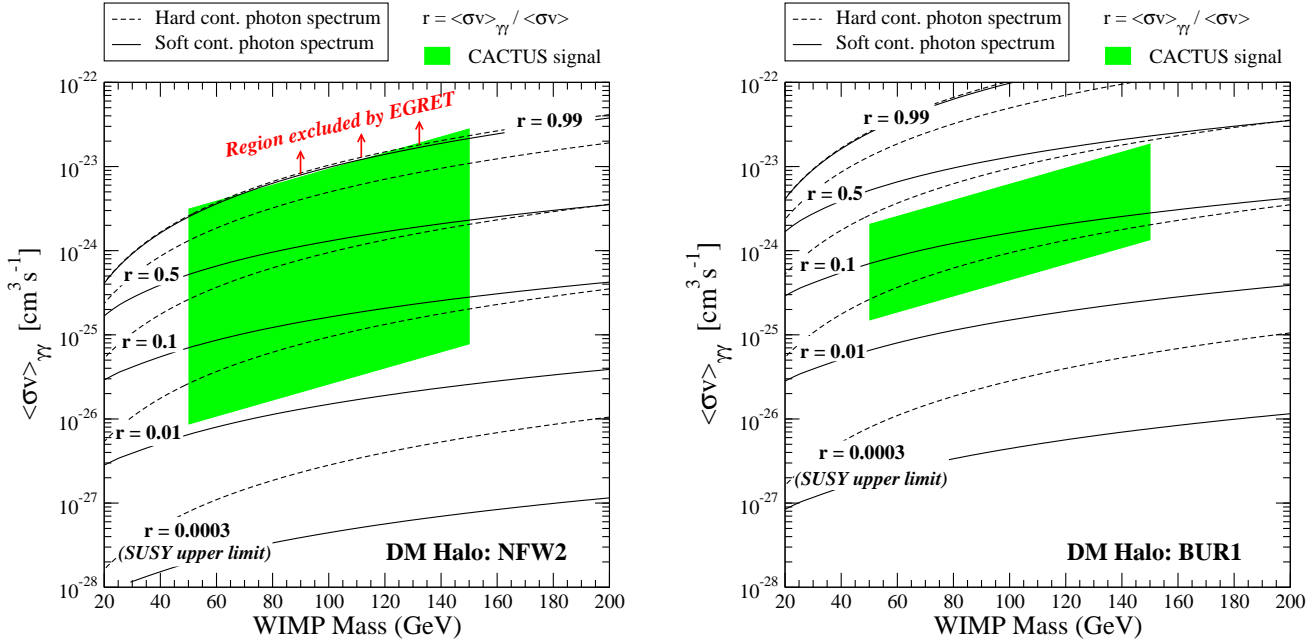

\begin{center}
\hspace*{-0.7cm}\mbox{\epsfig{file=plots/line_cu.eps,height=8.5cm}\quad\quad\epsfig{file=plots/line_co.eps,height=8.5cm}}
\end{center}
\caption{\it\small Limits in the $(m_\chi,\langle\sigma
v\rangle_{\gamma\gamma})$ plane to the ratio $r\equiv\langle\sigma
v\rangle_{\gamma\gamma}/\langle\sigma v\rangle_{\rm tot}$ from the
EGRET bound on the low-energy gamma-ray flux from Draco, for the
cuspy NFW2 profile (left) and for the cored BUR1 profile
(right), and for soft (solid lines) and hard (dashed lines)
continuum photon spectra.}
\label{fig:line}
\end{figure}

We assume, in Fig.~\ref{fig:line},
 hard and soft photon spectra (dashed and solid lines: the regions lying above the lines are excluded by the EGRET constraint),
and cuspy and cored profiles for the Draco dark-matter halo
(left and right panels). The green regions correspond to the
CACTUS signal: in the case of a
cored profile, the branching ratio into two photons must be at
least as large as 1$\div$10\%, clearly incompatible with
supersymmetric dark matter. Requiring that all of the
CACTUS excess counts originate from photons produced in
dark-matter annihilation, and consistency with the EGRET bound, implies much larger branching
fractions $(r\simeq 0.5-0.99)$ and a dark-matter model where the
dark-matter particle predominantly annihilates into two photons.

\subsection{Decaying dark matter}\label{sec:decay}

Uncertainties in the innermost structure of dark-matter halos
critically impact the
computation of the flux of photons from dark-matter
annihilation. Since the annihilation rate per unit volume is
proportional to the square of the dark-matter density,
the occurrence of high-density spikes or cusps in the center of
dark-matter halos may lead to flux enhancements of various orders
of magnitude, without affecting significantly the outer structure
of the dark halo, where the rotation curves are usually best
determined. If, alternatively, the gamma-ray excess results
from the {\em decay} of a quasi-stable dark-matter
particle, the flux is only linearly proportional to the dark matter density (and hence, in the case of a distant object like Draco, to a first approximation, to the total dark-matter
mass), which is much more reliably constrained by dynamical
measurements, inducing a signficantly smaller spread in the flux predictions.

We now show that the CACTUS Draco gamma-ray excess {\em
cannot} be explained in terms of a decaying dark-matter particle.
The resulting flux of photons in the diffuse gamma-ray background and from the center of our own Galaxy
would in fact be exceedingly large.
The ratio of the gamma-ray flux  $\phi^{\rm Draco}$ from Draco's
halo and that $\phi^{\rm MW}$ from the Milky
Way halo can be written as
\begin{equation}\label{eq:decay}
     r_{\rm GC,\ DB}\equiv\frac{\phi^{\rm Draco}}{\phi^{\rm
     MW}_{\rm GC,\ DB}}\simeq\left(\frac{M_{\rm\sss
     DM}^{\rm Draco}}{(d^{\rm Draco})^2}\right)\left(\int_{\rm
     line\ of\ sight} \rho^{\rm MW}(l){\rm d}l\right)^{-1},
\end{equation}
where the line-of-sight integral is performed either in the
direction of the Galactic center ($r_{\rm GC}$) or in the
direction of Draco, to evaluate the diffuse gamma-ray background
($r_{\rm DB}$). If the Draco mass is formed entirely
by dark matter, then $M_{\rm\sss DM}^{\rm
Draco}\simeq8.6\times10^7 M_\odot$  \cite{klenya1}. The distance
to Draco is
$d^{\rm Draco}\simeq76\ {\rm kpc}$ \cite{vdb}. The Milky Way
line-of-sight integrals depend on the assumed dark-matter halo.
We evaluated the integrals for two extreme choices, a NFW
profile \cite{nfw} with scaling density and radius
$\rho_s=5.4\times 10^6\ {M_\odot/{\rm kpc}^3}$ and $r_s=21.8\
{\rm kpc}$ respectively, and a Burkert profile \cite{bur}, with
scaling density and radius $\rho_s=1.5\times 10^7\ {M_\odot/{\rm
kpc}^3}$ and $r_s=11.7\ {\rm kpc}$. We find
\begin{equation}
     r_{\rm DB}\lesssim 1.5\div1.7\times 10^{-4}, \qquad r_{\rm
     GC}\lesssim 0.25\div68\times 10^{-6},
\end{equation}
with the smaller figures corresponding to the NFW Milky Way
profile. Over a 1 degree angular region, the diffuse gamma-ray
background from the direction of Draco quoted in
Eq.~(\ref{eq:diff}) gives an integrated flux over 50 GeV
$\phi^{50}_{\rm DB}\lesssim 10^{-11}\ \units$. We do not
have a direct measurement of the gamma-ray spectrum from the
Galactic center in the energy window between 10 GeV and 150 GeV,
the lower limit being the highest energy probed by EGRET, and the
upper limit the lowest energy probed by HESS. To have an idea of
the integral flux above 50 GeV one can extrapolate the power-law
behaviors featured by both the EGRET (at $E\gtrsim2$ GeV) and HESS
differential photon flux, respectively, giving
\begin{eqnarray}
     \nonumber &&\frac{{\rm d}\phi_{\rm\sss EGRET}}{{\rm
     d}E}\simeq1.6\times 10^{-6}\left(\frac{E}{\rm
     GeV}\right)^{-3.1}\ \diffunits, \\ 
     \nonumber &&\frac{{\rm d}\phi_{\rm\sss HESS}}{{\rm
     d}E}\simeq2.5\times 10^{-12}\left(\frac{E}{\rm
     TeV}\right)^{-2.2}\ {\rm cm}^{-2}{\rm s}^{-1}{\rm TeV}^{-1}. 
\end{eqnarray}
These figures, extrapolated and integrated over the energy range
$E>50\ {\rm GeV}$, give the following integrated fluxes:
\begin{eqnarray}
     \nonumber &&\phi^{50}_{\rm\sss EGRET}\simeq2\times
     10^{-10}\ \units, \\ 
     \nonumber &&\phi^{50}_{\rm\sss HESS}\simeq8\times 10^{-11}\
     \units.
\end{eqnarray}
The flux of gamma-ray photons from dark-matter decay from Draco
is therefore conservatively bounded to be
\begin{eqnarray}
     \phi^{50}_{\rm Draco}&\lesssim& 1.7\times
     10^{-15}\units,\qquad {\rm diffuse\ gamma-ray\
     background},\\ 
     \phi^{50}_{\rm Draco}&\lesssim& 1.4\times 10^{-14}\units,\qquad
     {\rm Galactic\ center},
\end{eqnarray}
which is clearly incompatible with the estimates in
Eq.~(\ref{eq:dracosig}).

We therefore conclude that the expected flux from the Galactic
center and from the diffuse gamma-ray background are not
consistent with the putative CACTUS signal from Draco in the
context of a decaying dark-matter particle. A sufficiently large
Draco flux would evidently violate the available
constraints to the diffuse gamma-ray background and on the flux
of gamma rays from the center of the Milky Way by various orders
of magnitude.

\section{Conclusions}\label{sec:conclusions}

We considered in this paper the possibility that the gamma-ray
excess observed by CACTUS from the direction of the Draco dSph
galaxy originates from WIMP annihilation. We
summarize below the main results of the present analysis:
\begin{itemize}

\item We showed that future measurements, with $\sim0.1$ degree
angular resolution, should allow us to distinguish between cored
and cuspy halos, even though the current CACTUS angular
resolution is not good enough.
\item We estimated the putative gamma-ray flux from dark-matter
annihilation considering the two extreme possibilities that the
dark-matter signal consists of (1) all the excess counts
reported by CACTUS, and (2) the excess over background
corresponding to Draco's innermost region.
\item We analyzed, in a model-independent approach, the regions
of the particle-mass versus annihilation-cross-section parameter
space compatible with the estimated CACTUS excess,
imposing the constraints from the null results of gamma-ray
searches reported by EGRET and by the VERITAS collaboration observation with the Whipple-10m ACT, and from the absence of a
statistically relevant excess in the CACTUS data for photon
energies above 150 GeV.  The annihilation cross section range lies well above that expected with standard thermal relic dark matter production.
\item The total excess counts over background
reported by CACTUS can only be reproduced with a very cored
dark-matter halo and a hard photon spectrum.  If this is the
correct explanation for the excess, then
higher-angular-resolution measurements should still show a
$\sim1$ degree spread around Draco's center.  Almost all
CACTUS-compatible models will be within reach of GLAST.
\item In the case of supersymmetric dark matter, the
annihilation cross sections needed to reproduce the CACTUS
signal are close to the maximal theoretically allowed values and would imply a negligible thermal relic abundance.
CACTUS-compatible supersymmetric models give typically very
large detection rates at direct-detection experiments, and a
sizeable neutrino flux from neutralino annihilation in the Sun,
which in principle could allow a cross-check of the gamma-ray
signal that GLAST should see within this scenario.
\item The possibility of a dominantly monochromatic origin of
the CACTUS excess is not viable within supersymmetry, and
requires, for consistency with the EGRET null result from Draco,
dark-matter models that annihilate predominantly to two photons
and thus produce a very suppressed continuum photon spectrum.
\item A decaying-dark-matter scenario is ruled out by the
resulting inferred gamma-ray flux from the center of the Milky
Way and in the diffuse gamma-ray background.
\end{itemize}


\vspace*{1cm}
\noindent{ {\bf Acknowledgments} } \\
\noindent We would like to thank Max Chertok, Francesc Ferrer, Jeter Hall, Reshmi Mukherjee,
Mani Tripathi, and Piero Ullio for helpful
discussions.  This work was supported in part by DoE
DE-FG03-92-ER40701 and FG02-05ER41361 and NASA NNG05GF69G.



\begin{thebibliography}{99}
\small

\bibitem{dmreviews1}
  G.~Jungman, M.~Kamionkowski and K.~Griest,
  Phys.\ Rept.\  {\bf 267} (1996) 195
  [arXiv:hep-ph/9506380].
\bibitem{dmreviews2}
   L.~Bergstr\"om,
  Rept.\ Prog.\ Phys.\  {\bf 63}, 793 (2000)
  [arXiv:hep-ph/0002126];   G.~Bertone, D.~Hooper and J.~Silk,
  Phys.\ Rept.\  {\bf 405}, 279 (2005)
  [arXiv:hep-ph/0404175].

\bibitem{glast}
  W.~Atwood {\it et al.}  [GLAST-LAT Collaboration],
{\it Prepared for 11th International Conference on Calorimetry in
High-Energy Physics (Calor 2004), Perugia, Italy, 28 Mar - 2 Apr
2004}

\bibitem{ACTreview}
  F.~A.~Aharonian and C.~W.~Akerlof,
  Ann.\ Rev.\ Nucl.\ Part.\ Sci.\  {\bf 47} (1997) 273;
  F.~Aharonian,
  arXiv:astro-ph/0511139;

\bibitem{pierobuck}
  L.~Bergstr\"om, P.~Ullio and J.~H.~Buckley,
  Astropart.\ Phys.\  {\bf 9}, 137 (1998)
  [arXiv:astro-ph/9712318].

\bibitem{deboer}
  W.~de Boer, C.~Sander, A.~V.~Gladyshev and D.~I.~Kazakov,
  arXiv:astro-ph/0508617.

\bibitem{galcenter}
  D.~Hooper, I.~de la Calle Perez, J.~Silk, F.~Ferrer and S.~Sarkar,
  JCAP {\bf 0409}, 002 (2004)
  [arXiv:astro-ph/0404205];
  D.~Horns,
  Phys.\ Lett.\ B {\bf 607}, 225 (2005)
  [Erratum-ibid.\ B {\bf 611}, 297 (2005)]
  [arXiv:astro-ph/0408192];
  D.~Hooper and J.~March-Russell,
  Phys.\ Lett.\ B {\bf 608}, 17 (2005)
  [arXiv:hep-ph/0412048].

\bibitem{profumo}
  S.~Profumo,
  Phys.\ Rev.\ D {\bf 72} (2005) 103521
  [arXiv:astro-ph/0508628].

\bibitem{tyler}
  C.~Tyler,
  Phys.\ Rev.\ D {\bf 66} (2002) 023509
  [arXiv:astro-ph/0203242].


\bibitem{baltz}
  E.~A.~Baltz, C.~Briot, P.~Salati, R.~Taillet and J.~Silk,
  Phys.\ Rev.\ D {\bf 61}, 023514 (2000)
  [arXiv:astro-ph/9909112];

\bibitem{pieri}
  L.~Pieri and E.~Branchini,
  Phys.\ Rev.\ D {\bf 69}, 043512 (2004)
  [arXiv:astro-ph/0307209];
  N.~Fornengo, L.~Pieri and S.~Scopel,
  Phys.\ Rev.\ D {\bf 70}, 103529 (2004)
  [arXiv:hep-ph/0407342].

\bibitem{extragal}
    D.~Elsaesser and K.~Mannheim,
  Phys.\ Rev.\ Lett.\  {\bf 94}, 171302 (2005)
  [arXiv:astro-ph/0405235];

\bibitem{evans}
  N.~W.~Evans, F.~Ferrer and S.~Sarkar,
  Phys.\ Rev.\ D {\bf 69}, 123501 (2004)
  [arXiv:astro-ph/0311145].


\bibitem{dsphrev}
M.L.~Mateo, ARA\&A, {\bf 36}, 435 (1998).

\bibitem{dracoradio}
E.B.~Fomalont and B.J.~Geldzahler, Astron. J. {\bf 84}, 12 (1979).

\bibitem{cactus}
P.~Marleau, TAUP, Zaragoza, Spain, September 2005;  M.~Tripathi,
Cosmic Rays to Colliders 2005, Prague, Czech Republic, September
2005; TeV Particle Astrophysics Workshop, Batavia, USA, July 2005;
M.~Chertok, proceedings of PANIC 05, Santa Fe, USA, October 2005.

\bibitem{nogammatyler}
L.M.~Young, Astron. J. {\bf 117}, 1758 (1999).

\bibitem{hooper}
L.~Bergstr\"om and D.~Hooper, arXiv:astro-ph/0512317.

\bibitem{mash}
S.~Mashchenko, A.~Sills and H.M.P.~Couchman,
arXiv:astro-ph/0511567.

\bibitem{nfw}
  J.~F.~Navarro, C.~S.~Frenk and S.~D.~M.~White,
  Astroph. J. {\bf 499}, L5 (1998) [arXiv:astro-ph/9611107].

\bibitem{aaronson}
M.~Aaronson, Astroph. J. {\bf 266}, L11 (1983).

\bibitem{aaronole}
M.~Aaronson and E.W.~Olszewski (1987) in ``Dark Matter in the
Universe'' (IAU Symposium No.117), eds. J.~Kormendy, G.R.~Knapp,
(Reidel, Dordrecht), p.153; M.~Aaronson and E.W.~Olszewski (1988)
in ``Large Scale Structure of the Universe'' (IAU Symposium
No.130), eds. J.~Audouze, M.C.~Pelletan, A.~Szalay (Kluwer,
Dordrecht), p.153.

\bibitem{armand}
T.E.~Armandroff, E.W.~Olszewski and C.~Prior AJ {\bf 110}, 2131
(1995).

\bibitem{hargre}
J.C.~Hargreaves, G.~Gilmore, M.J.~Irwin and D.~Carter, MNRAS {\bf
282}, 305 (1996).

\bibitem{klenya1} J.~Klenya {\em et al.}, Ap.~J. 563 (2001) L115.


\bibitem{klenya2}
J.~Klenya {\it et al.}, MNRAS {\bf 330}, 792 (2002).

\bibitem{wilkinson2004}
M.I.~Wilkinson {\it et al.}, ApJ {\bf 611}, L21 (2004).

\bibitem{kroupa}
P.~Kroupa, New Astronomy {\bf 2}, 139 (1997).

\bibitem{bur}
  A.~Burkert, Astrophys. J. {\bf 447} (1995) L25;  P.~Salucci and A.~Burkert, Astrophys. J. {\bf 537} (2000) L9.

\bibitem{egret} {\tt http://cossc.gsfc.nasa.gov/egret/}

\bibitem{wai}
L.~Wai, ``{\em Analysis of Draco with EGRET}'',\\ {\tt  http://www-glast.slac.stanford.edu/ScienceWorkingGroups/DarkMatter/oldstuff/9-9-02.ppt}

\bibitem{lamb} R.~C.~Lamb and D.~J.~Macomb, Ap.~J. 488 (1997) 872.

\bibitem{Morselli:2002nw}
  A.~Morselli, A.~Lionetto, A.~Cesarini, F.~Fucito and P.~Ullio  [GLAST
                  Collaboration],
  Nucl.\ Phys.\ Proc.\ Suppl.\  {\bf 113} (2002) 213
  [arXiv:astro-ph/0211327].

\bibitem{jeter} 
J.~Hall, private communication, 2005, and   J.~Hall  [VERITAS Collaboration],
  arXiv:astro-ph/0507448. {\tt http://veritas.sao.arizona.edu/}.

\bibitem{nonth}
B.~Murakami and J.~D.~Wells,
Phys.\ Rev.\ D {\bf 64}, 015001 (2001);
T.~Moroi and L.~Randall,
Nucl.\ Phys.\ B {\bf 570}, 455 (2000);
M.~Fujii and K.~Hamaguchi,
Phys.\ Lett.\ B {\bf 525}, 143 (2002);
M.~Fujii and K.~Hamaguchi,
Phys.\ Rev.\ D {\bf 66}, 083501 (2002);
R.~Jeannerot, X.~Zhang and R.~H.~Brandenberger,
JHEP {\bf 9912}, 003 (1999);
W.~B.~Lin, D.~H.~Huang, X.~Zhang and R.~H.~Brandenberger,
Phys.\ Rev.\ Lett.\  {\bf 86}, 954 (2001);

\bibitem{enh}
M.~Kamionkowski and M.~S.~Turner,
Phys.\ Rev.\ D {\bf 42} (1990) 3310;
%
P.~Salati,
[arXiv:astro-ph/0207396];
F.~Rosati,
Phys.\ Lett.\ B {\bf 570} (2003) 5
[arXiv:hep-ph/0302159];
%
S.~Profumo and P.~Ullio,
JCAP {\bf 0311}, 006 (2003)
[arXiv:hep-ph/0309220];
%
R.~Catena, N.~Fornengo, A.~Masiero, M.~Pietroni and F.~Rosati,
arXiv:astro-ph/0403614;
%
S.~Profumo and P.~Ullio,
Proceedings of the {\em 39th Rencontres de Moriond Workshop on
Exploring the Universe: Contents and Structures of the Universe},
La Thuile, Italy, 28 Mar - 4 Apr 2004, ed. by J. Tran Thanh Van [arXiv:hep-ph/0305040].
%

\bibitem{egretgc}
  H.A.~Mayer-Hasselwander et al., Astron. Astrophys. {\bf 335}
  (1998) 161.


\bibitem{Profumo:2004ty}
  S.~Profumo and P.~Ullio,
  JCAP {\bf 0407} (2004) 006
  [arXiv:hep-ph/0406018].

\bibitem{Baer:2005tw}
  H.~Baer and S.~Profumo,
  JCAP {\bf 0512} (2005) 008
  [arXiv:astro-ph/0510722].


\bibitem{vdb}
  S.~Van den Bergh, ``{\em The Galaxies of the Local Group}'', Cambridge, UK: Cambridge University Press, 2000.

\bibitem{Spergel:2003cb}
  D.~N.~Spergel {\it et al.}  [WMAP Collaboration],
  Astrophys.\ J.\ Suppl.\  {\bf 148} (2003) 175
  [arXiv:astro-ph/0302209].

\end{thebibliography}
\end{document}